\documentclass[twocolumn,aps,groupedaddress,superscriptaddress,amsmath,amssymb]{revtex4-2}
\usepackage[utf8]{inputenc}
\usepackage{enumerate}
\usepackage{amsmath}
\usepackage{braket}
\usepackage[margin=1in]{geometry}
\usepackage{float}
\usepackage{graphicx}
\usepackage{siunitx}
\usepackage{upgreek}
\usepackage[version=4]{mhchem}
\usepackage[normalem]{ulem}
\usepackage{xcolor}
\usepackage{fancyhdr}
\usepackage{natbib}
\begin{document}



\author{S.~Norimoto}
\affiliation{National Physical Laboratory, Hampton Road, Teddington TW11 0LW, United Kingdom}
\author{R.~Saxena}
\affiliation{National Physical Laboratory, Hampton Road, Teddington TW11 0LW, United Kingdom}
\author{P.~See}
\affiliation{National Physical Laboratory, Hampton Road, Teddington TW11 0LW, United Kingdom}
\author{A.~Nasir}
\affiliation{National Physical Laboratory, Hampton Road, Teddington TW11 0LW, United Kingdom}
\author{J.~P.~Griffiths}
\affiliation{Cavendish Laboratory, University of Cambridge, J. J. Thomson Avenue, Cambridge CB3 0HE, United Kingdom}
\author{C.~Chen}
\affiliation{Cavendish Laboratory, University of Cambridge, J. J. Thomson Avenue, Cambridge CB3 0HE, United Kingdom}
\author{D.~A.~Ritchie}
\affiliation{Cavendish Laboratory, University of Cambridge, J. J. Thomson Avenue, Cambridge CB3 0HE, United Kingdom}
\author{M.~Kataoka}
\affiliation{National Physical Laboratory, Hampton Road, Teddington TW11 0LW, United Kingdom}

\title{Photon emission by hot electron injection across a lateral \textit{pn} junction}

\begin{abstract}
We demonstrate a method to generate photons by injecting hot electrons into a {\it pn} junction within a \ce{GaAs/AlGaAs} heterostructure.
Hot electrons are generated by biasing across a mesoscopic potential in {\it n}-type region and travel toward {\it p}-type region through quantum Hall edge channel in the presence of magnetic field perpendicular to the substrate.
The {\it p}-type region is created several microns away from the hot electron emitter by inducing interfacial charges using a surface gate.
The energy relaxation of the hot electrons is suppressed by separating the orbitals before and after longitudinal-optical (LO) phonon emission.
This technique enables the hot electrons to reach the {\it p}-type region and to recombine with induced holes followed by photon emissions.
Hot electron-induced hole recombination is confirmed by a peak around \qty{810}{nm} in an optical spectrum that corresponds to excitonic recombination in a \ce{GaAs} quantum well.
An asymmetric structure observed in the optical spectrum as a function of the magnetic field originates from the chiral transport of the hot electrons in the Hall edge channel.
We propose the combination of our technology and on-demand single-electron source would enable the development of an on-demand single photon source that is an essential building block to drive an optical quantum circuit and to transfer quantum information for a long distance.
\end{abstract}

\date{\today}

\maketitle

\section{Introduction}
Ideal single photon sources for the field of quantum optics towards applications in quantum communication, computing, and imaging would be on-demand~\cite{Senellart2017}.
Here, ``on-demand'' does not just mean that the device emits one photon when requested, but also means that the device does not emit any photons when not requested.
Such sources may require the controlled preparation of an excited state to avoid accidental emission of the second photon, either simultaneously or within a delay shorter than the detector jitter duration.  
There is usually a trade-off between the photon emission rate (or brightness) and single-photon purity characterised by the minimum coincidence counts $g^{(2)}(0)$~\cite{wang2016, Somaschi2016, Senellart2017, Michler2000}.
This issue can be mitigated to some extent, for example, by using an optical cavity to enhance coupling between a quantum dot and excitation laser, and a source with a brightness of 65\% and $g^{(2)}(0)$ less than 1\% has been observed~\cite{Somaschi2016}. 
However, as long as the emitter is pumped by continuous excitation (optically or electrically) albeit for a short duration in pulse, it is difficult to eliminate the probability of second photon emission completely.  

A different approach was proposed by Foden~{\it et~al.} in 2000~\cite{foden2000high}.
In their proposed scheme, electrons are carried one by one by surface acoustic waves~\cite{Shilton1996} across a lateral {\it n-i-p} junction that is biased below the threshold.
As long as the injection of electrons is separated long enough compared to the radiative recombination time, and as long as only one electron is injected at a time, the probability of second photon emission is, in principle, negligibly small.
A recent experiment~\cite{hsiao2020single} demonstrated a photon source exceeding the classical limit using this scheme, although $g^{(2)}(0)$ was as high as 39\%.  
This was most likely due to poor current quantisation in the experiment.  
The probability of injecting two electrons in one cycle was not eliminated.
Guaranteeing single-electron transfer in surface acoustic waves devices has been challenging~\cite{Chung2019} with the lowest error rate demonstrated so far being as high as $10^{-4}$ in some devices~\cite{Janssen2001}.
Tunable-barrier quantum dot pumps~\cite{Blumenthal2007}, on the other hand, can routinely achieve single-electron transfer with an error rate as low as $10^{-7}$~\cite{Stein2017}.
Blumenthal~{\it et~al.} suggested that such pumps could be used for a variable frequency photon emitter if the emitted electrons can be injected into a {\it pn} junction placed nearby~\cite{Blumenthal2007}.
Unlike surface acoustic wave-based single electron transfer, it is trivial to emit electrons at arbitrary intervals, therefore tunable-barrier pumps can be ideal for on-demand single-photon sources.
Buonacorsi~{\it et~al.} listed single-photon source as one of the potential applications from their single-electron pumps on dopant-free material~\cite{Buonacorsi2021}. 
Meanwhile, a method has been developed to create a lateral {\it pn} junction on an {\it n}-type substrate by converting a part of the substrate into a {\it p}-type region~\cite{Dobney2023}.
This allows a relatively straightforward integration of the existing tunable-barrier electron pump device architecture verified with the timing control of single-electron emission as accurate as one picosecond~\cite{Fletcher2013, Kataoka2016, Johnson2018, Fletcher2019}.

An on-demand single-photon source is required to suppress the emission of photons when not requested.  
The literature does not provide the details of how electrons could be injected into a {\it pn} junction to enable on-demand control without getting accidental photon emissions.  
In this paper, we first propose a scheme for on-demand single-photon emission that would suppress unwanted photon emission.  
In our scheme presented in Section I\hspace{-0.05pt}I, electrons from a tunable-barrier pump are injected into a {\it pn} junction as hot electrons with energy appreciably larger than the thermal energy.
The {\it pn} junction is biased below the threshold voltage so that the energy step at the junction prevents thermally excited electrons from entering the {\it p}-type region, while only hot electrons from the pump can enter the {\it p}-type region.
In Section I\hspace{-0.05pt}I\hspace{-0.05pt}I, we demonstrate this effect, revealing the electroluminescence by hot-electrons from a continuous source injected into a lateral {\it pn} junction.
As the hot-electron source, we use a mesoscopic potential barrier with a bias across it.
Emitted hot electrons are transported to a {\it pn} junction through high-energy edge states~\cite{Fletcher2013} with an edge-depletion gate to suppress inelastic scattering process~\cite{Kataoka2016, Johnson2018}.
The electroluminescence spectrum indicates that the photon emission process happens when hot electrons enter the {\it pn} junction.

\section{Proposed single-photon source driven by tunable-barrier single-electron pump}
\begin{figure*}[!htb]
    \centering
    \includegraphics[width=15cm]{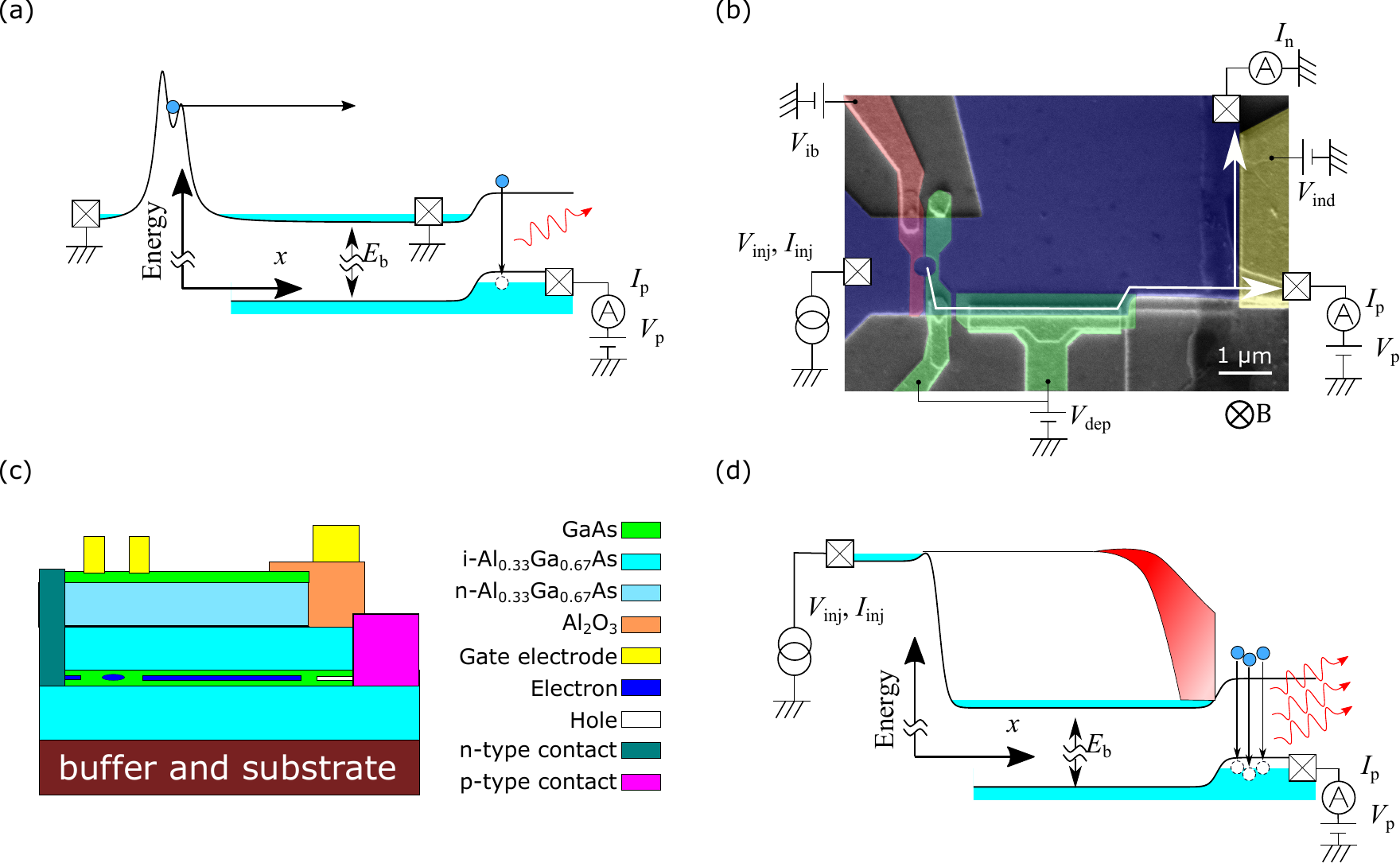}
    \caption{(a)~Schematic for an on-demand single-photon source using a tunable-barrier tunnel quantum dot pump. A hot electron is emitted from a tunable-barrier single-electron source and enters into a \textit{p}-type region followed by recombination with a hole resulting in the emission of a single-photon.  (b)~A scanning electron microscope image of the device in false colour and circuit diagram. Ohmic contacts are indicated with a cross mark in a box and magnetic field $B$ is applied perpendicular to the substrate. Each region is indicated with a false colour. (Blue) Two dimensional electron gas system. (Red) Injection barrier gate to control the energy of hot electrons. (Green) Depletion gate to control LO phonon emission rate for hot electrons. (Yellow) Inducing gate to induce holes underneath. Hot electron trajectory at a negative high magnetic field is indicated with white arrows. (c)~Schematic cross-section of the device (d)~Energy diagram along with the hot electron trajectory. Applying negative voltage on $V_{\mathrm{dep}}$ modifies the optical phonon emission rate, which enables the transport of hot electrons for long distances with maintaining their energy. Hot electrons that overcome a potential step at the \textit{pn} junction enter into the \textit{p}-type conduction band and recombine with holes in the valence band followed by photon emissions.}
    \label{Schematics}
\end{figure*}
Figure~\ref{Schematics}~(a) shows a schematic of the energy band structure for our proposed on-demand single-photon source using a tunable-barrier single-electron pump.
The junction is biased below the threshold for a forward current so that electrons/holes in the Fermi sea (including those thermally excited) cannot travel across the junction (to eliminate the background photon emission).
A tunable-barrier quantum-dot pump is placed near the {\it pn} junction.
An electron emitted by the quantum-dot pump needs to travel as a hot electron until it enters the {\it p}-type region.
Since this is the only electron in the conduction band, when it radiatively recombines with a hole in the valence band, only one photon matching the band-gap energy is emitted.
As a single-electron source is capable of emitting a hot electron on demand\cite{Fletcher2013, Fletcher2019, Fletcher2023, Ubbelohde2023}, a single-photon is expected to be emitted on-demand.

A crucial point in this scheme is to maintain the hot electron energy while the electron travels from the pump to the {\it p}-type region.
We propose to guide hot electrons into a quasi-one-dimensional channel formed by quantum Hall edge states in the presence of a magnetic field applied perpendicular to the plane of the two-dimensional electron gas (2DEG).
In order to suppress the accidental charging by hot electrons losing their energy and thermalising onto the Fermi surface before reaching the {\it pn} junction, the intermediate 2DEG in the region between the source and {\it pn} junction is connected to the device ground to sink these electrons.

The main inelastic scattering process for electrons travelling in the high-energy quantum Hall edge states in a GaAs system is found to be longitudinal-optical (LO) phonon emissions~\cite{Kataoka2016, Johnson2018, Emary2016, Emary2019, Ota2019, Taubert2011}.
When the hot electron loses enough energy, its trajectory is blocked by the potential step at the \textit{pn} junction.
Modifying the edge potential profile to a shallower slope by surface gate voltage can separate hot-electron trajectories before and after LO phonon emission, reducing the scattering probability~\cite{Kataoka2016, Johnson2018, Emary2019}.
This enables a transfer of a hot electron over a long distance keeping its energy.
Our proposed device structure includes such an edge depletion gate to enable hot-electron transport from the source to {\it p}-type region.

While the previous experimental demonstrations in the literature listed above showed hot-electron transport from a single-electron pump to a mesoscopic barrier acting as an electron detector, it has not yet been demonstrated whether hot-electron transport length can be extended into a {\it p}-type region.
Complications may arise as the application of a negative gate voltage on the edge-depletion gate may attract holes and hence may cause a leakage current from the {\it p}-type region. 
In the following sections, we describe our recent experimental results showing that such a single-photon source is feasible by demonstrating electroluminescence caused by the injection of hot electrons emitted from a continuous source.

\section{Experimental Setup}
In order to demonstrate that a hot electron can be transported from a mesoscopic device to a {\it p}-type region and recombine with a hole to emit a photon, we fabricated a \textit{pn} junction integrated with a hot-electron source.
Our device is fabricated on a conventional 2DEG system of \ce{GaAs/AlGaAs} quantum-well heterostructure with electron density and mobility \qty{1.5e15}{\m^{-2}} and \qty{100}{m^{2}/Vs} at \qty{1.5}{K}, respectively.
The device consists of an \textit{n}-type region from the original wafer structure on the substrate and a \textit{p}-type region where the {\it n}-type dopant layer has been etched away and holes are induced by gate voltage operation~\cite{Dobney2023}.
Figure~\ref{Schematics}~(b) shows a scanning electron microscope image and circuit diagram.
The \textit{n}-type region indicated in blue is fabricated by conventional photolithography and electron-beam lithography techniques, wet mesa etching, metal film deposition with vacuum evaporator and alloying layers of \ce{Ge/Au/Ni} to form electric contacts to the 2DEG.
Gate electrodes of \ce{Ti/Au} are deposited by a vacuum evaporator.

A voltage $V_{\mathrm{ib}}$ on the injection barrier gate indicated in red in Fig. \ref{Schematics}~(b) creates a mesoscopic barrier.
Applying a voltage $V_{\mathrm{inj}}$ across the barrier drives a current $I_{\mathrm{inj}}$, injecting hot electrons with energy at around $-eV_{\mathrm{inj}}$.
Applying gate voltage $V_{\mathrm{dep}}$ on the depletion gates indicated in green makes the edge potential profile shallower and suppresses LO phonon emissions under the gates so that hot electrons maintain their energy to overcome a potential step at the {\it pn} junction~\cite{Kataoka2016, Johnson2018}.
Holes are induced under the inducing gate indicated in yellow by an applied gate voltage $V_{\mathrm{ind}}$.
We follow the method detailed in Ref.~\cite{Dobney2023} to create a {\it p}-type region on an {\it n}-type wafer.
Figure~\ref{Schematics}~(c) shows a schematic of the cross section to provide an idea how to make a \textit{pn} junction on a conventional 2DEG wafer.
First, the dopant layer, \textit{n}-\ce{AlGaAs}, is etched away to eliminate the 2DEG in this region.
Second, a layer of \ce{Au/Be} is deposited and annealed to create {\it p}-type ohmic contacts as the sources of holes.
Then, a layer of \ce{Al2O3} is deposited by the atomic layer deposition technique, followed by the deposition of a \ce{Au/Ti} gate layer.

Figure~\ref{Schematics}~(d) shows an energy band diagram with schematic circuits along the bottom edge in Fig.~\ref{Schematics}~(b) when a strong magnetic field is applied on the device (into the paper, for which we set the value as negative).
Hot electrons are injected through the potential barrier created by the injection barrier gate~\cite{Fujisawa2019}.
A constant DC current $I_{\mathrm{inj}}$ is injected through the mesoscopic barrier from the contact on the left using a source measure unit (SMU).
The bias voltage $V_{\mathrm{inj}}$ applied across the barrier is monitored by the same SMU.
We adopted this method of controlling the constant $I_{\mathrm{inj}}$ rather than controlling the constant $V_{\mathrm{inj}}$, because if a large injection current flows with a constant $V_{\mathrm{inj}}$, it causes a positive feedback loop between the resultant large amount of photon emission and the lowering of the potential barrier due to the ionisation of dopants by photon absorption, leading to an even larger injection current.
If the current level is kept low enough with a constant $I_{\mathrm{inj}}$, the rate of dopant ionisation is kept at a negligible level and the barrier height could be kept constant during the measurements.

The hot electrons emitted from the mesoscopic barrier travel along the bottom edge as shown by arrows in Fig.~\ref{Schematics}~(b).
At the boundary of \textit{n}-type and \textit{p}-type region, the electrons that have maintained enough energy to overcome the \textit{pn} junction potential barrier step will enter the p-type region.
These electrons recombine with holes and emit photons.
The photons are captured by a lens assembly mounted on a tri-axial piezo stage and are guided into a spectrometer through an optical fibre.
A SMU connected to \textit{p}-type contact detects the current $I_{\mathrm{p}}$ (the current flowing through the {\it pn} junction) that contributes to photon emission.
Electrons reflected at the potential barrier step flow into the top ohmic contact and the current $I_{\mathrm{n}}$ is detected by a digital multimeter (DMM) via a transimpedance amplifier.

The amount of $I_{\mathrm{p}}$ depends on voltages applied on the depletion gate (green) $V_{\mathrm{dep}}$ and \textit{p}-type contact $V_{\mathrm{p}}$.
$V_{\mathrm{dep}}$ modifies the sample edge potential and suppresses LO phonon emission rate for hot electrons by separating the orbital before and after LO phonon scattering, which allows hot electrons to survive for a long distance\cite{Johnson2018, Fletcher2019}.
As band gap is intrinsically defined by the material, $V_{\mathrm{p}}$ modifies the potential step height at the \textit{pn} junction.
If the step height is lowered enough, electrons in the Fermi sea can enter the \textit{p}-type region.
The thresholds for forward bias current are $V_{\mathrm{p}} =$ \qty{1.525}{V} and \qty{1.535}{V} at $B = \qty{0}{T}$ and \qty{-10}{T}, respectively.
In order to overcome the charging effect that affects the density of holes, we periodically switch the inducing gate voltage $V_{\mathrm{ind}}$ to refresh the accumulated charge most likely at the insulator-semiconductor interface~\cite{Dobney2023}.
$V_{\mathrm{ind}}$ is set at \qty{+5}{V} for \qty{2}{sec} followed by switching to the value of $V_{\mathrm{p}}$ (so that the gate voltage is zero relatives to the holes potential)\cite{Dobney2023}.
The switching of $V_{\mathrm{ind}}$, controlled by GS200, triggers the spectrometer to acquire spectrum for \qty{2}{sec}.
That also triggers the DMM and SMUs to record currents and voltages as a function of time for \qty{3}{sec}. 
\section{Result}
\begin{figure}[!hbtp]
    \centering
    \includegraphics[width=7.9cm]{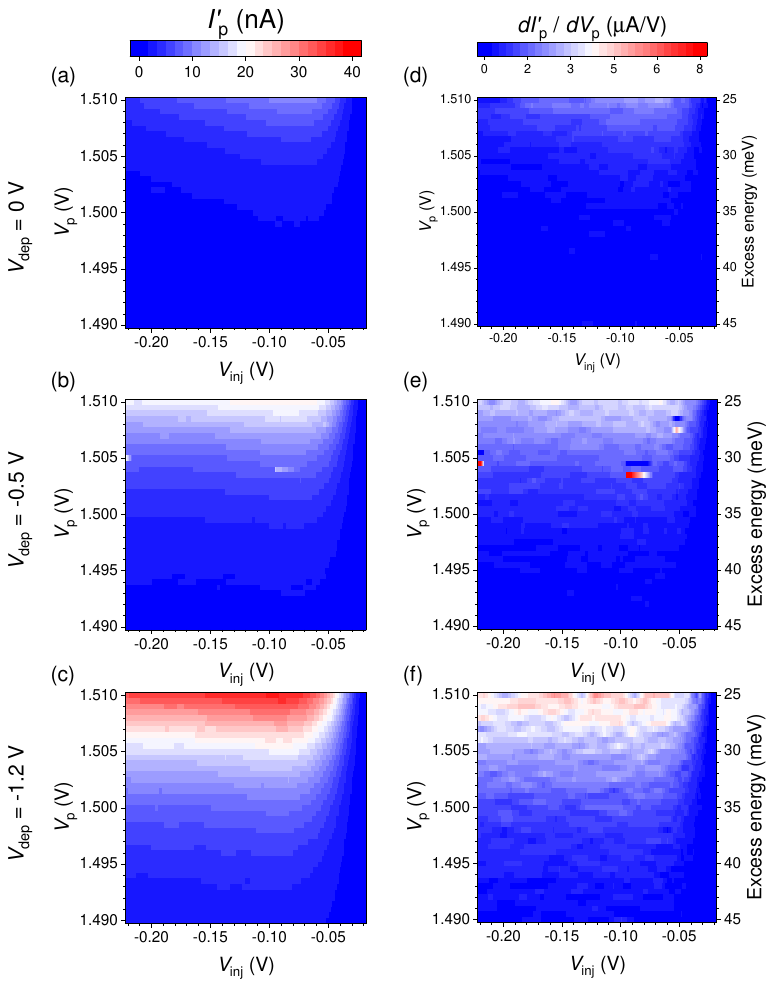}
    \caption{Left: Colour plots of excess current though \textit{p}-type contact $I'_{\mathrm{p}}$ as functions of $V_{\mathrm{p}}$ and $V_{\mathrm{inj}}$ at $V_{\mathrm{dep}} = $ (a)~\qty{0}{\V}, (b)~\qty{-0.5}{\V} and (c)~\qty{-1.2}{\V}. They share colour scale shown above them. Right: Derivatives of t3he left-hand side plots with respect to $V_{\mathrm{p}}$ at $V_{\mathrm{dep}} = $ (d)~\qty{0}{\V}, (e)~\qty{-0.5}{\V} and (f)~\qty{-1.2}{\V}. They are also plotted in the same scale shown above them. All data here are taken at $V_{\mathrm{p}} = \qty{1.510}{V}$, $I_{\mathrm{inj}} =$ \qty{300}{\nA} and $B = \qty{-10}{\tesla}$. The scales on the right-hand side axis indicate excess energy from Fermi surface in the \textit{n}-type region.}
    \label{HotElectronInjection}
\end{figure}
Figure~\ref{HotElectronInjection} shows excess current $I'_{\mathrm{p}}=I_{\mathrm{p}}(t=\qty{50}{ms})-I_{\mathrm{p}}(t=\qty{3}{s})$ 
as functions of $V_{\mathrm{inj}}$ and $V_{\mathrm{p}}$ at $V_{\mathrm{dep}} =$ (a)~\qty{0}{\V}, (b)~\qty{-0.5}{\V} and (c)~\qty{-1.2}{\V} on the left panel, and their derivative with respect to $V_{\mathrm{p}}$ on the right panel (d-f).
Figure~\ref{HotElectronInjection}~(a-c) are computed with interpolation based on excess current $I'_{\mathrm{p}}$ and applied bias voltage on hot electron reservoir $V_{\mathrm{inj}}$ as functions of  $V_{\mathrm{ib}}$ and $V_{\mathrm{p}}$ at $I_{\mathrm{inj}} = \qty{300}{\nA}$ and $B = \qty{-10}{\tesla}$.
As the step height of the \textit{pn} junction potential barrier is modified by $V_{\mathrm{p}}$, ~\ref{HotElectronInjection}~(d)-(f) reflect hot electron energy distribution at the respective $V_{\mathrm{dep}}$.
The energy scale of hot electrons from Fermi level is indicated on the right-hand side axis of each graph. 
The zero-current region on the top-right corner of each plot clearly indicates that a finite injection bias $V_{\mathrm{inj}}$ is needed to transfer hot electrons into the \textit{p-}type region.
When the maximum energy of hot electrons is smaller than \textit{pn} junction barrier step height, no electrons can enter into the \textit{p}-type region as shown in the region where $V_{\mathrm{inj}}$ is less negative than \qty{-0.03}{V} (i.e. $V_{\mathrm{inj}}>\qty{-0.03}{V}$).
As further evidence that hot electrons enter into the \textit{p}-type region, the amount of current $I'_{\mathrm{p}}$ is enhanced by applying negative voltage on $V_{\mathrm{dep}}$.
The majority of hot electrons are expected to lose their energy on the trajectory between the deletion gate and the inducing gate where optical phonon emission is not suppressed (we deliberately left this gap due to the alignment tolerance margin in the device fabrication).
It is also found that there is a threshold for $V_{\mathrm{inj}}$ to control energy distribution.
In the range of around $V_{\mathrm{inj}}= [-0.07,\qty{-0.03}{V}]$, the maximum energy of injected electrons is proportional to the injection voltage.
~In the regime that $V_{\mathrm{inj}}$ is more negative than around \qty{-0.07}{V}, the energy distribution is insensitive to $V_{\mathrm{inj}}$.
\begin{figure}[!htb]
    \centering
    \includegraphics[width=7.9cm]{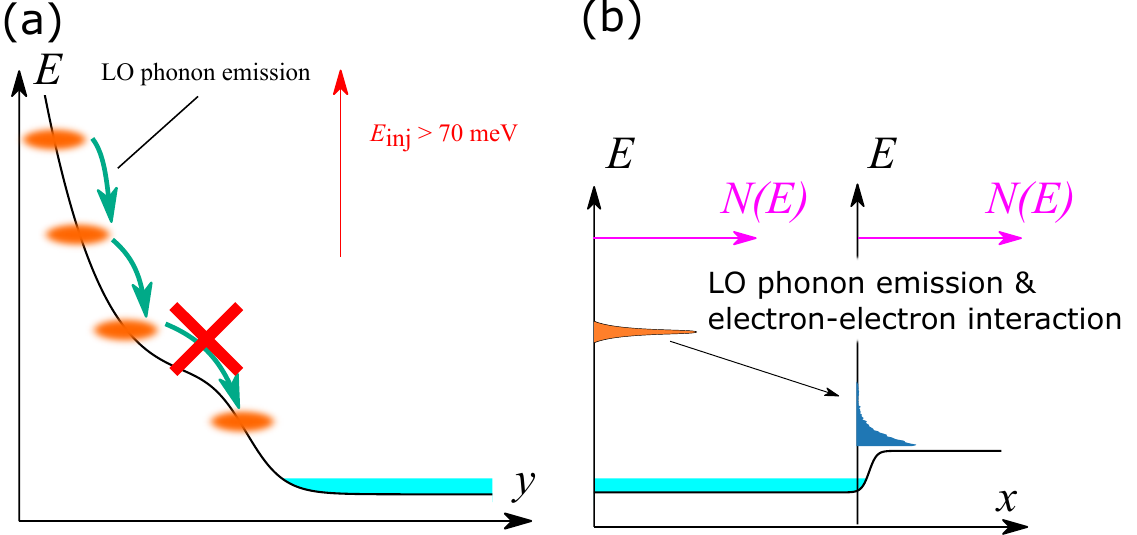}
    \caption{(a)~A Landau level around 2DEG edge to show energy relaxation process under the depletion gate. The energy level is indicated in the y coordinate perpendicular to propagation. An orange circle represents a hot electron distribution. In the regime $E_{\mathrm{inj}}$ is greater than \qty{70}{meV}, energy distribution at the end of the depleted area is initialised by certain energy distribution due to LO phonon emission process. (b)~The figure shows hot electrons' energy distributions at the end of the depleted area and the \textit{pn} junction.  Hot electrons (orange) relax into lower energy distribution (blue) on the way to \textit{pn} junction via LO phonon emission and electron-electron interaction. The energy distribution at the \textit{pn} junction is an average in the range of $V_{\mathrm{inj}}=[-0.22, \qty{-0.10}{V}]$ in Fig. \ref{HotElectronInjection}~(f).}
    \label{relaxation}
\end{figure}
It is expected that the LO photon emission suppression technique is available for hot electrons within a certain energy window.  
Figure~\ref{relaxation}~(a) shows a Landau level with respect to the direction perpendicular to the sample edge, where a hot electron propagates.
The step in the Landau level is developed by the depletion gate electrode.
The dominant relaxation process for a hot electron is LO phonon emission.
The phonon emission rate for a hot electron is suppressed when its orbital is separated from the orbital of the lower energy state in space due to Fermi's golden rule~\cite{Johnson2018, Emary2016, Emary2019}.
Therefore, a hot electron with higher energy than the potential step loses its energy even under the depletion gate as there are large overlaps in its orbital before and after photon emission.
Injected electrons in the high energy regime keep relaxing into the lower energy state until they reach the state with low phonon emission rate, which leads to a constant energy distribution against $V_{\mathrm{inj}}$ at the end of the depleted area.
Figure~\ref{relaxation}~(b) shows energy relaxation for the constant energy distribution between the depleted area and \textit{pn} junction.
The energy distribution at the end of the depletion gate is indicated by orange filling as a function of energy.
The energy distribution at \textit{pn} junction indicated by blue filling is computed by taking an average in the range of $V_{\mathrm{inj}}=[-0.22, \qty{-0.10}{V}]$ in Fig. \ref{HotElectronInjection}~(f).
As the phonon emission process is not suppressed for a few microns between the depletion gate and \textit{pn} junction, hot electrons relax into lower energy states immediately via LO phonon emission at high energy regime and electron-electron interaction around Fermi surface.

The chiral edge transport of hot electrons is also observed in the electroluminescence spectrum under the magnetic field.
\begin{figure}[!htb]
    \centering
    \includegraphics[width=7.9cm]{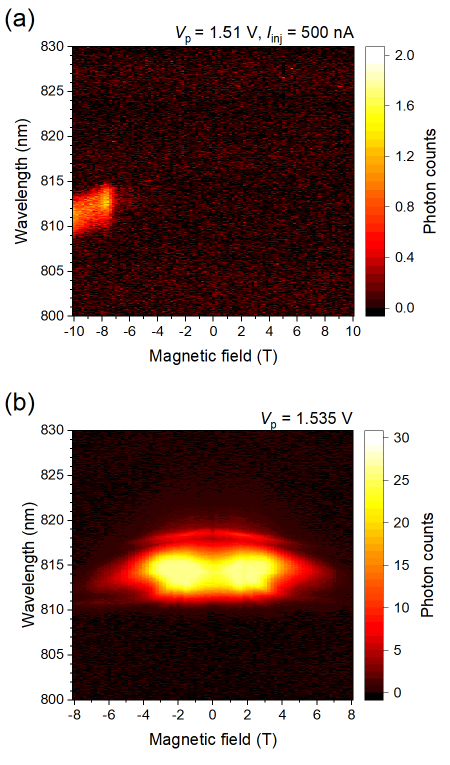}
    \caption{(a)~Magnetic field dependence of photon emission spectrum with hot electrons at $I_{\mathrm{inj}} =$ \qty{500}{\nA} and $V_{\mathrm{p}} = \qty{1.510}{\V}$. (b)~Magnetic field dependence of photon emission spectrum with electrons in Fermi sea at $V_{\mathrm{p}} = \qty{1.535}{\V}$. 2 (a) and (b) are measured at the same piezo stage position.}
    \label{Symmetry}
\end{figure}
Figure~\ref{Symmetry} shows magnetic field dependence of photon emission spectrum from the \textit{pn} junction caused by (a)~hot electrons and (b)~electrons in the Fermi sea at the same piezo stage position, which provides evidence that injected hot electrons travelling quantum Hall edge channel are converted into photons.
A spectrum at each magnetic field consists of an averaged spectrum with an offset to make the background signal level equal to zero.
Raw spectra are filtered to eliminate strong peak signals generated by cosmic rays and are averaged 30 times to improve noise-signal ratio.

Hot electrons of \qty{500}{nA} are injected at $V_{\mathrm{p}} = \qty{1.510}{V}$ that is too small for electrons in the Fermi sea to overcome the \textit{pn} junction barrier step.
In Fig.~\ref{Symmetry}~(a), electroluminescence is observed only at high negative magnetic fields where injected hot electrons can enter into the p-type region due to quantum Hall edge channel.
At the lower fields, hot electrons cannot maintain their energy with enhanced phonon emissions~\cite{Johnson2018, Emary2019}, and therefore cannot reach the {\it pn} junction.
At the positive fields, the electrons emitted from the injector follow the opposite chirality and do not reach the {\it pn} junction. 
On the other hand, in Fig.~\ref{Symmetry}~(b), large enough $V_{\mathrm{p}}$ is applied to produce electroluminescence without hot electron injection.
The symmetric pattern with respect to the magnetic field implies that the edge-state chirality is not important for the electroluminescence by the Fermi sea, as electrons/holes can enter the {\it pn} junction at both edges.
These results show that a hot electron travels through quantum Hall edge channel and overcomes a potential barrier step at \textit{pn} junction followed by photon emission with recombination. 

\section{Conclusion}
A proof of concept that hot electrons can be converted into photons is successfully demonstrated using hot electrons travelling along a quantum Hall edge channel.
Hot electron energy spectroscopy using a potential step at the {\it pn} junction shows monotonic energy distributions as a function of injection energy and a constant hot electron distribution at high injection energy regime.
The observation suggests that hot electron distribution is initialised under the depletion gate and that the distribution relaxes on the way to the {\it pn} junction due to LO phonon emission and electron-electron interaction.
Hot electrons with energies greater than the potential step enter the {\it p}-type region followed by photon emission.
An asymmetric optical spectrum from the {\it pn} junction with respect to the magnetic field tells that chiral transport in quantum Hall edge channel plays an important role in transferring hot electrons into the {\it p}-type region.
In principle, a single hot electron injection from a tunable-barrier tunneling electron source would generate a single photon with \textit{pn} junction free from second photon emission.

\section{Acknowledgement}
This project has received funding from the European Union's H2020 research and innovation programme under grant agreement No 862683.

\section{Appendix: Discussion in electron-photon conversion rate}
Conversion efficiency from electron to photon, often called quantum efficiency, is one important parameter in an optoelectronic or electro-optic device.
We estimate the electron-photon conversion rate including photon collection efficiency of the lens unit, fibre couplings, fibre loss and loss at each optical component.
The number of electrons injected into \textit{p}-type region and the number of photons detected at the spectrometer are estimated by fitting current through \textit{p}-type contact $I_{\mathrm{p}} $ as a function of time with exponential and fitting spectrum with Gaussian, respectively.
These fittings indicate that \qty{1e9} electrons are required to detect one photon at the spectrometer.
If tunable-barrier tunnelling single-electron source works at an operation frequency of \qty{1}{GHz} and all pumped electrons enter the \textit{p}-type region, the spectrometer captures only one photon in \qty{1}{sec}.
The over-focusing lens unit and not optimised fibre position are considered major contributions to reduce collection efficiency.
Improvement in collection efficiency is required to conduct coincidence count $g^{(2)}(0)$ measurement in a realistic timescale.
\bibliographystyle{junsrt}
\bibliography{bibliography}
\end{document}